\documentclass[sigconf]{acmart}

\usepackage{booktabs}
\usepackage{graphicx}
\usepackage{amsmath}
\usepackage{amssymb}
\usepackage{multirow}
\usepackage{array}
\usepackage{xcolor}

\setcopyright{none}
\renewcommand\footnotetextcopyrightpermission[1]{}
\newcommand{\gsid}{\textsc{GSID}}
\newcommand{\method}{TS-GEAR}

\title{\method: Taobao Shangou GEnerative Ad Retrieval}

\newcommand{\AlibabaAffiliation}{%
  \affiliation{%
    \institution{Alibaba Group}
    \city{Shanghai}
    \country{China}%
  }%
}
\author{Guofeng Tong}
\AlibabaAffiliation
\email{guofengtong.tgf@alibaba-inc.com}
\author{Xu Zhang}
\AlibabaAffiliation
\email{shoucai.zx@alibaba-inc.com}
\author{Guangliang Yu}
\AlibabaAffiliation
\email{yuguangliang.ygl@alibaba-inc.com}
\author{Haiqiang Wang}
\AlibabaAffiliation
\email{wang.whq@alibaba-inc.com}
\author{Guannan Cong}
\AlibabaAffiliation
\email{congguannan.cgn@alibaba-inc.com}
\author{Xing Guo}
\AlibabaAffiliation
\email{danghu.gx@alibaba-inc.com}
\author{Yang Zhou}
\AlibabaAffiliation
\email{liaowu.zy@alibaba-inc.com}
\author{Jingjing Duan}
\AlibabaAffiliation
\email{duan.djj@alibaba-inc.com}
\author{Shuang Li}
\AlibabaAffiliation
\email{reubenli.ls@alibaba-inc.com}
\author{Feng Shi}
\AlibabaAffiliation
\email{sam.sf@alibaba-inc.com}
\author{Hualei Liu}
\AlibabaAffiliation
\email{jianglang@taobao.com}

\newcommand{\TSGearAuthorNames}{Guofeng Tong, Xu Zhang, Guangliang Yu,
  Haiqiang Wang, Guannan Cong, Xing Guo, Yang Zhou, Jingjing Duan,
  Shuang Li, Feng Shi, Hualei Liu}
\hypersetup{pdfauthor={\TSGearAuthorNames}}
\authorsaddresses{}

\newsavebox{\TSGearAuthorLineBox}
\newcommand{\TSGearFitAuthorLine}[1]{%
  \sbox{\TSGearAuthorLineBox}{#1}%
  \ifdim\wd\TSGearAuthorLineBox>\textwidth
    \resizebox{\textwidth}{!}{\usebox{\TSGearAuthorLineBox}}%
  \else
    \usebox{\TSGearAuthorLineBox}%
  \fi
}
\makeatletter
\def\@mkauthors@iii{%
  \global\setbox\mktitle@bx=\vbox{%
    \unvbox\mktitle@bx\par\medskip
    \centering
    {\fontsize{10}{12}\selectfont
      \TSGearFitAuthorLine{Guofeng Tong, Xu Zhang,
        Guangliang Yu, Haiqiang Wang, Guannan Cong, Xing Guo}\par
      \TSGearFitAuthorLine{Yang Zhou, Jingjing Duan, Shuang Li,
        Feng Shi, Hualei Liu}\par}%
    {\normalsize Alibaba Group\par}%
    {\normalsize Shanghai, China\par}%
    {\fontsize{8}{10}\selectfont
      \TSGearFitAuthorLine{\{guofengtong.tgf,shoucai.zx,yuguangliang.ygl,wang.whq,%
        congguannan.cgn,danghu.gx\}@alibaba-inc.com}\par
      \TSGearFitAuthorLine{\{liaowu.zy,duan.djj,reubenli.ls,sam.sf\}%
        @alibaba-inc.com,\,jianglang@taobao.com}\par}%
    \bigskip
  }%
}
\AtBeginDocument{%
  \let\TSGearOriginalMaketitle\maketitle
  \renewcommand{\maketitle}{%
    \TSGearOriginalMaketitle
    \hypersetup{pdfauthor={\TSGearAuthorNames}}%
    \ifdefined\XeTeXversion
    \fi
  }%
}
\makeatother

\acmConference[WSDM '27]{The 20th ACM International Conference on Web Search and
  Data Mining}{February 15--19, 2027}{Hong Kong, China}
\acmYear{2027}
\copyrightyear{2027}

\begin{document}

\begin{abstract}
On-demand delivery platforms such as DoorDash and Uber Eats impose a
constraint absent in traditional e-commerce: merchants beyond the user's
delivery range are \emph{invalid} candidates rather than merely low-ranked.
Generative retrieval has recently shown strong potential in large-scale
retrieval systems. Adapting it to this setting, however, poses three
challenges. First, existing semantic IDs encode geography only as a soft
signal, so reachability can only rely on a non-differentiable post-generation
hard truncation. Second, user preferences vary substantially across both
regions and queries, so aggregating historical behaviors without conditioning
on geography and the current query yields biased interest estimates. Third,
the retrieved candidate set is evaluated by its collective value, yet
autoregressive training scores each sequence independently, with no set-level
signal to discourage redundant near-duplicates. To address these challenges,
we propose \textbf{\method}, a unified generative retrieval framework for
on-demand delivery: (1) \textbf{Geo-Semantic ID} splits merchant identifiers
into geographic prefixes and semantic suffixes, making reachability a
prediction target rather than a post-hoc filter; (2) \mbox{\textbf{Query-Geo Target
Attention}} aggregates historical behaviors using the real-time query and
location as dual targets, while encoding each behavior's own location into its
attention key; (3) \textbf{GMPO} (Group Marginal Policy Optimization) rewrites advantages using a deduplicated group
reward and a Marginal Advantage within GRPO, avoiding auxiliary diversity
objectives. Deployed online with a lightweight 0.1B backbone, \method{} serves
over 150 million queries per day in the search advertising pipeline of
Taobao Shangou, consistently outperforming state-of-the-art generative retrieval
methods on both HitRate and geographic validity, and achieving 3.8\% gains in
ad impressions and 4.2\% in ad revenue.
\end{abstract}

\keywords{generative retrieval, reinforcement learning, on-demand delivery,
search advertising}

\maketitle

\section{Introduction}
Taobao Shangou is an on-demand delivery platform that connects users with
nearby merchants through a local logistics network, offering services such as
food delivery and instant retail (e.g., groceries and electronics).
Specifically, the merchants available to a user are limited to those within
delivery range of the user's real-time location, and orders are typically
fulfilled by a courier within an hour. Reachability is thus a hard constraint in
Taobao Shangou search: the system must not only understand the user's query
intent but also assess, in real time, whether a merchant can deliver to the
user. This fundamentally distinguishes it from general e-commerce search.

\begin{figure}[t]
  \centering
  \includegraphics[width=\columnwidth]{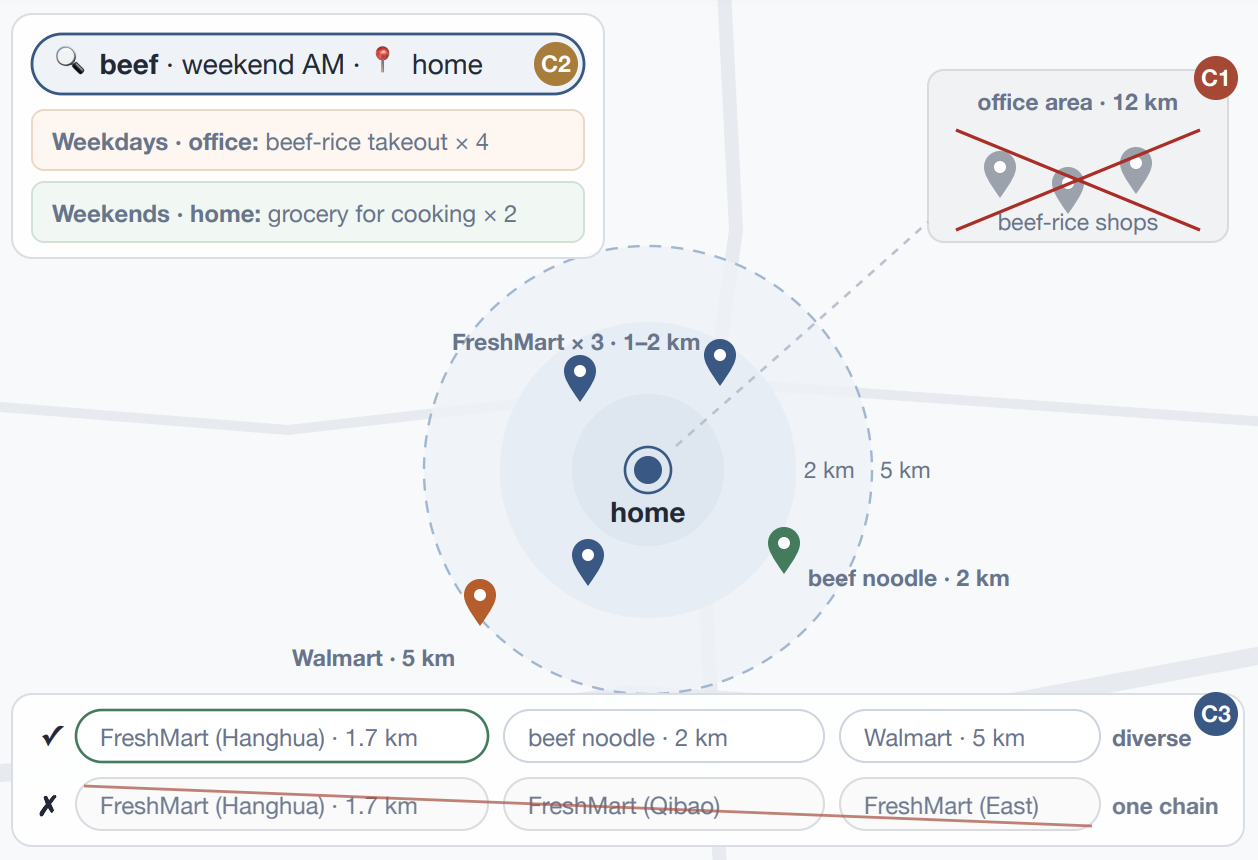}
  \caption{A motivating request in on-demand delivery search. The same query
  ``beef'' maps to hot ready-to-eat dishes near the office on weekdays, but to
  fresh ingredients at home on weekends, illustrating the three challenges.}
  \label{fig:intro}
\end{figure}

In recent years, retrieval has been shifting from the traditional two-tower
paradigm to generative retrieval. This new paradigm represents each item as a
semantic ID (SID) and directly generates candidate SIDs with an autoregressive
model, naturally unifying multi-source signals (query semantics, user behavior,
etc.) under a single objective. It can further be aligned to business
objectives via reinforcement learning, and has been validated in industry by
works such as TIGER~\cite{tiger}, OneRec~\cite{onerec}, and OneSug~\cite{onesug}.
However, deploying generative retrieval in Taobao Shangou's sponsored search
poses three key challenges. We illustrate them with a concrete example
(Figure~\ref{fig:intro}): user A frequently orders food delivery near his office
on weekdays, with his order history dominated by hot-food shops (e.g., beef
rice). On weekends he typically cooks at home. One weekend morning, at home, he
opens Taobao Shangou and searches ``beef''.

\textbf{Challenge 1: geographic reachability is not structurally enforced.}
Existing SIDs are derived solely from content or collaborative signals,
so high-probability candidates are biased toward the beef-rice shops near user
A's office, semantically relevant yet beyond the home delivery range. Such
candidates can only be removed by a non-differentiable post-generation filter,
which wastes beam budget and lies outside end-to-end optimization.

\textbf{Challenge 2: interest bias in behavior aggregation.} For the same query
``beef'', the user's intent is ready-to-eat dishes at the office but fresh
ingredients at home. When behaviors are aggregated without geographic
conditioning, the representation is dominated by the more frequent
office-vicinity preference, causing the system to retrieve hot-food merchants
instead of the fresh-grocery ones the user actually wants. To model user
interest accurately, the system must condition on the real-time query--location
pair and account for the geographic context in which each past behavior
occurred.

\textbf{Challenge 3: mismatch between token-level training and set-level
objective.} Autoregressive training scores each sequence independently, so the
beam is easily filled by near-duplicate merchants from a single fresh-grocery
chain, each correct on its own but redundant as a set, crowding out
alternatives such as beef rolls or noodles. Retrieval is essentially
\emph{next-set prediction}, whose quality lies in the whole set's cumulative
value and diversity; token-level likelihood also cannot differentiate strong
signals (e.g., orders) from weak ones (e.g., clicks).

To address these challenges, we propose \method{}, a generative retrieval
framework for on-demand delivery, built on an encoder--decoder
architecture and optimized through supervised pre-training followed by RL
alignment. We introduce three designs, each targeting one
challenge: (1) the \textbf{Geo-Semantic ID} splits a merchant identifier into a
geographic prefix and a semantic suffix, so that a prefix-level constrained
decoding turns reachability from a post-filter into a structural generation
target, while a \textbf{User-Geo Gate} inside the decoder injects the user's real-time
location to condition the order in which reachable grids are generated; (2) \mbox{\textbf{Query-Geo Target
Attention}} aggregates behaviors with the real-time query and location as dual
targets, enabling location-aware interest modeling and cross-region behavior
transfer; (3) \textbf{GMPO} (Group Marginal Policy Optimization) reformulates
the GRPO advantage with a deduplicated group reward and a Marginal Advantage,
aligning set-level value without any auxiliary diversity objective.

Our \textbf{contributions} are summarized as follows:
\begin{itemize}
  \item We propose a unified generative retrieval framework designed for
  on-demand delivery, unifying geographic hard-constraint modeling and
  set-level value alignment within a single encoder--decoder architecture.
  \item We design the Geo-Semantic ID that, via a prefix-level constrained
  decoding, moves geographic reachability from a post-filter into the generation
  process, and incorporate a User-Geo Gate inside the decoder that injects the user's
  real-time location to condition the order of the generated grids.
  \item We propose Query-Geo Target Attention, which jointly conditions on
  geographic context and query semantics in the encoder, enabling cross-region
  user interest modeling.
  \item We propose the GMPO alignment algorithm, which reformulates GRPO's
  advantage with a deduplicated group reward and a leave-one-out Marginal
  Advantage, directly optimizing the cumulative value and diversity of the
  retrieved set while suppressing policy collapse.
  \item Experiments on industrial and public datasets show
  that our method consistently outperforms state-of-the-art generative retrieval
  methods on HitRate and geographic validity. In a 7-day online A/B test, the
  framework achieves +3.8\% ad impressions and +4.2\% ad revenue.
\end{itemize}

\section{Related Work}

\subsection{Generative Retrieval}
Generative retrieval reformulates the traditional two-tower ``encode + ANN''
pipeline as a sequence generation task that directly decodes item identifiers,
circumventing the representational bottleneck of a single inner product and the
encoding--indexing mismatch~\cite{dssm,youtubednn}. The paradigm was established
for document retrieval by DSI~\cite{dsi} and NCI~\cite{nci}; TIGER~\cite{tiger}
introduced RQ-VAE~\cite{rqvae} residual quantization, built on the discrete
codebooks of VQ-VAE~\cite{vqvae}, to obtain SIDs whose hierarchical
codebook structure enables coarse-to-fine generation, and the approach was
scaled to production by OneRec~\cite{onerec}, OneSug~\cite{onesug}, and
OneSearch~\cite{onesearch}. TSGR~\cite{tsgr} further introduced value-aware
SIDs with query-conditioned token orderings, while LLM-based variants
such as P5~\cite{p5} and LC-Rec~\cite{lcrec} leverage pre-trained language
models to unify recommendation tasks as prompted text generation.
HierGR~\cite{hiergr}, targeting food-delivery search, uses an LLM backbone but
only performs offline inference with cached reuse, modeling neither the user's
real-time location nor the behavior history. All these works derive identifiers
solely from content or collaborative signals, and their tokens encode no
explicit spatial structure, so geographic reachability can only be imposed as a
post-filter on already generated candidates, discarding a large fraction of the
beam in location-based service (LBS) settings and precluding end-to-end optimization.

\subsection{Geo-Aware Recommendation}
Spatial-context modeling has a long history in next-POI tasks, evolving from
early Markov models and metric embeddings to RNN- and attention/graph-based
methods such as DeepMove~\cite{deepmove}, STAN~\cite{stan}, and
GETNext~\cite{getnext}, and more recently to LLM-based next-POI models such as
LLM4POI~\cite{llm4poi}; their commonality is injecting distance or regional
features into the scoring function. Mainstream sequential recommenders such as
GRU4Rec~\cite{gru4rec}, SASRec~\cite{sasrec}, and BERT4Rec~\cite{bert4rec}
encode interaction histories effectively but are location-agnostic. Recent work
has begun to introduce geographic information into the generative paradigm:
GNPR~\cite{gnprsid} couples geographic and semantic signals within RQ-VAE
quantization, and OneLoc~\cite{oneloc} biases decoding toward nearby merchants
via geographic soft guidance. However, both model geography as a soft signal
(fused features or decoding bias), which can encourage proximity but cannot
structurally enforce reachability; on-demand delivery is stricter:
merchants beyond the courier's delivery range are invalid results. Moreover, these
methods operate in a query-free recommendation setting, and their public POI
datasets lack query signals, so geo-aware modeling has rarely been studied in a
search setting, leaving the joint treatment of query intent and spatial
reachability largely open.

\subsection{Reinforcement Learning Alignment}
\begin{figure*}[t]
  \centering
  \includegraphics[width=\linewidth]{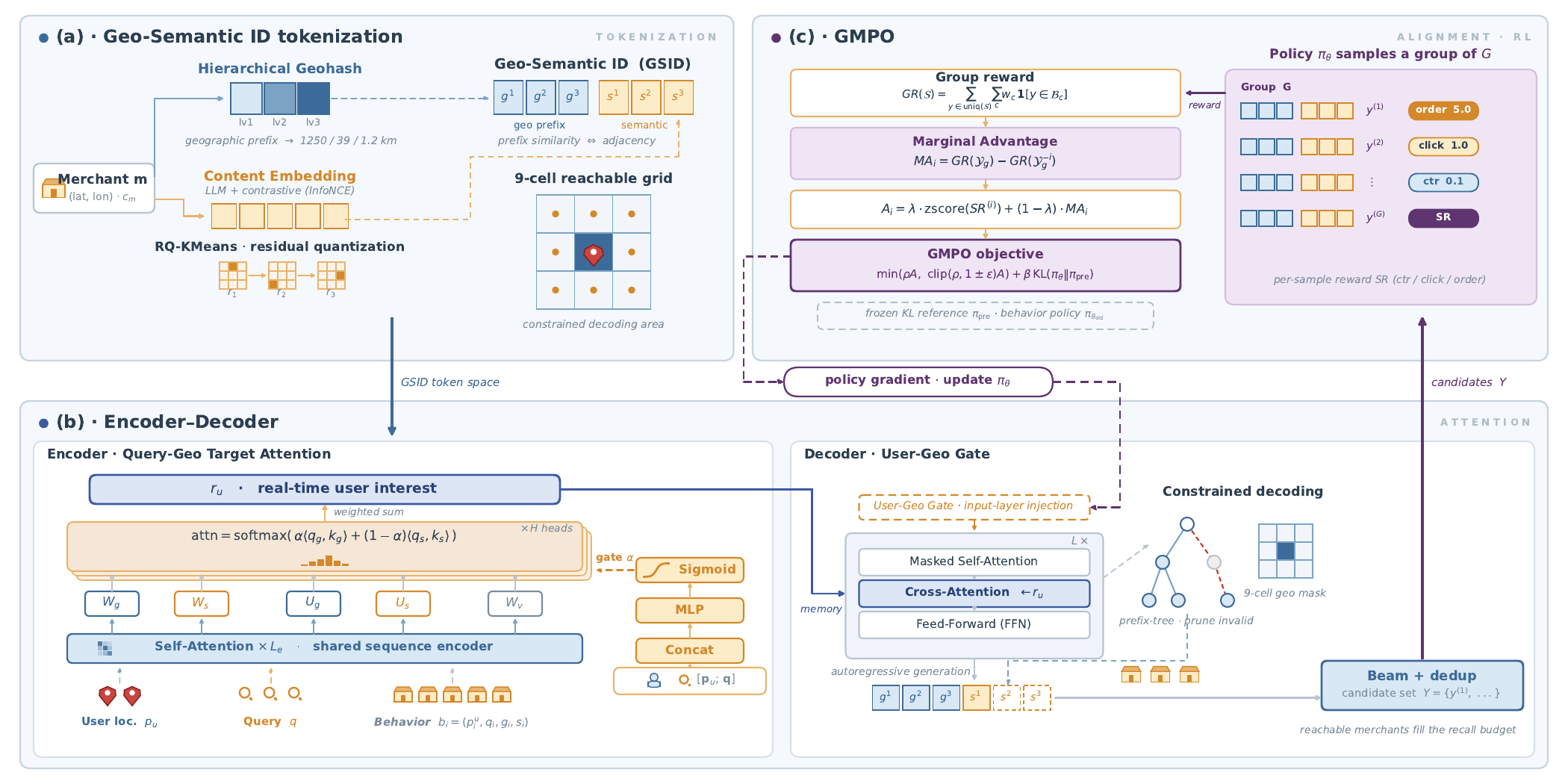}
  \caption{Overall architecture of \method{}. (a) Geo-Semantic ID
  tokenization, which builds hierarchical merchant identifiers. (b) The
  geo-aware search-interest encoder--decoder: a Query-Geo Target Attention
  encoder paired with a geo-constrained decoder that applies the User-Geo Gate.
  (c) GMPO alignment of the decoder policy.}
  \label{fig:arch}
\end{figure*}
Since the introduction of RLHF~\cite{rlhf}, aligning generative models with
downstream objectives via reinforcement learning has become standard practice.
PPO~\cite{ppo} established the RL pipeline; subsequent variants progressively
reduce training cost: the critic-free GRPO~\cite{grpo}, preference-based DPO~\cite{dpo},
and its reference-free variant SimPO~\cite{simpo}. These techniques have
been adapted to generative retrieval: OneSug~\cite{onesug} applies DPO with
hierarchical behavior rewards; OneSearch-V2~\cite{onesearchv2} refines GRPO with
a Token-Position Marginal Advantage that assigns position-aware credit to each
SID token, respecting the coarse-to-fine causal structure;
GFlowGR~\cite{gflowgr} applies GFlowNet to decompose item-level reward into
token-wise flow supervision, training generation probability proportional to
terminal reward. However, all these methods reward each sequence independently,
ignoring that retrieval delivers a \emph{set}: two relevant but near-duplicate
merchants contribute far less than their sum. Single-sequence optimization thus
drives the beam to collapse into a narrow region of the item space.

\section{Method}

\subsection{Problem Formulation}
\label{sec:prob}
Let the user set be $\mathcal{U}$ and the merchant set be $\mathcal{M}$. Each
merchant $m\in\mathcal{M}$ has a geographic location
$(\text{lat}_m,\text{lon}_m)$ and content features
$\mathbf{c}_m\in\mathbb{R}^{d_c}$ (e.g., images and region). A search request
is characterized by a triple $(q,\,p_u,\,\mathcal{B})$, where $q$ is the query
text, $p_u$ is the user's real-time location, and
$\mathcal{B}=\{b_i\}_{i=1}^{L_b}$ is the user's behavior history. Each behavior
$b_i=(p_i^{u},\,q_i,\,g_i,\,s_i)$ records the user's location $p_i^{u}$, the
query $q_i$ (a default placeholder in non-search contexts), and $g_i,s_i$
identifying the location and semantic attributes of the target merchant.
Throughout, $p$-symbols denote user-side locations and $g$-symbols merchant-side
geographic codes, and boldface (e.g., $\mathbf{p}_u$) denotes the corresponding
embedding.

We represent each merchant as a Geo-Semantic ID (\gsid) sequence
$\mathbf{y}=[g^{(1)},g^{(2)},g^{(3)},s^{(1)},s^{(2)},s^{(3)}]$, with three
geographic tokens followed by three semantic tokens. Retrieval is then a
constrained set-generation problem: to learn a policy $\pi_\theta$
that generates a deduplicated candidate set
$\mathcal{Y}=\{\mathbf{y}^{(1)},\dots,\mathbf{y}^{(B)}\}$ maximizing the whole
set's cumulative business value, subject to the hard constraint that the
geographic tokens of every candidate fall in the user-centric delivery range
$\mathcal{N}_u$:
\begin{equation}
\begin{split}
  &\max_{\theta}\;\mathbb{E}_{\mathcal{Y}\sim\pi_\theta(\cdot\mid q,p_u,\mathcal{B})}
  \big[\text{Value}(\mathcal{Y})\big] \\
  &\quad\text{s.t.}\;\;(g^{(1)},g^{(2)},g^{(3)})\in\mathcal{N}_u,\;\forall\,\mathbf{y}\in\mathcal{Y}.
\end{split}
\end{equation}
Our retrieval model aims to align with this objective; other online modules
enforce the hard constraints.
Unlike the token-wise likelihood maximization of language modeling,
$\text{Value}(\mathcal{Y})$ is defined at the candidate-set level; this
set-level property motivates our subsequent alignment design.

\subsection{Overall Architecture}
Our method follows an encoder--decoder architecture with a three-stage design.
The first stage, Geo-Semantic ID tokenization, builds a hierarchical merchant
representation tailored to on-demand delivery, decoupling geographic
location from content semantics. The second stage, autoregressive pre-training,
trains the encoder--decoder with Query-Geo Target Attention and User-Geo Gate to
jointly model user interest and real-time geographic context. The third stage,
set-level alignment, aligns the generation policy with set-level business
objectives via GMPO. The overall architecture is shown in
Figure~\ref{fig:arch}.

\subsection{Geo-Semantic ID Tokenization}
\label{sec:gsid}
A merchant's delivery range is not a fixed radius but a dynamic region that
depends on real-time fulfillment capacity (courier supply, weather, distance);
reachability therefore cannot be precomputed as a static attribute. Meanwhile,
naively embedding geographic information into the SID cannot
structurally enforce reachability during generation. Our design rationale is to
discretize space into fine-grained geographic cells and let the model, at
generation time, locate the nearest cells that jointly cover the user's delivery
range, so that reachability is expressed as an explicit, decodable prefix rather
than a post-hoc filter. We accordingly introduce the Geo-Semantic ID, which
augments the SID with an explicit spatial hierarchy and adaptively
captures fulfillment constraints by generating a geographic prefix code.

\noindent\mbox{\textbf{Geographic prefix code.}} The geographic prefix code is obtained
via hierarchical geohash~\cite{geohash} encoding. Balancing spatial granularity
against codebook size for on-demand delivery, we select the 6-digit geohash
(geohash-6) as the finest geographic unit: for merchant $m$, we take its 6-digit
base-32 geohash (spanning roughly $1.2\,\text{km}\times0.6\,\text{km}$ grid) and
split it into three chunks of two characters each to obtain the hierarchical
geographic code:
\begin{equation}
g_m^{(k)} = \text{Chunk}_k\!\big(\text{Geohash}_6(\text{lat}_m,\text{lon}_m)\big).
\end{equation}
Each chunk $g_m^{(k)}$ consists of two base-32 characters, with a vocabulary of
size $|\mathcal{V}_g|=32\times32=1024$; as $k$ increases, the region encoded by
the first $2k$ characters narrows from roughly
$1250\,\text{km}\to 39\,\text{km}\to 1.2\,\text{km}$. Note that geohash cells are
not a strictly regular grid. Cell dimensions vary with latitude and the aspect
ratio changes across digit positions, so all grid sizes above are approximate.

\noindent\mbox{\textbf{Content semantic code.}} The content semantic code is obtained by
quantizing the merchant embedding with a residual quantizer. To specialize the
embedding space for on-demand delivery, we construct structured prompts
incorporating region, brand, category, and frequent queries, and fine-tune a
pre-trained Qwen3-Embedding~\cite{qwen3emb} model via contrastive learning with
in-domain positive/negative pairs. Following OneRec, we then apply RQ-KMeans for
multi-level residual quantization to produce the three-token content semantic
code:
\begin{equation}
s_m^{(k)} = \text{RQ-KMeans}_{k}(\mathbf{c}_m),\quad k\in\{1,2,3\}.
\end{equation}
Concatenating the two hierarchical sequences yields the Geo-Semantic ID:
\begin{equation}
\text{GSID}(m) = \big[\,g_m^{(1)},g_m^{(2)},g_m^{(3)},\,s_m^{(1)},s_m^{(2)},s_m^{(3)}\,\big]
\;\in\;\mathcal{V}_g^{\,3}\times\mathcal{V}_s^{\,3}.
\end{equation}
This representation has two properties. First, \emph{prefix similarity implies
spatial proximity}: geohash preserves lexicographic locality, so merchants
sharing a geographic prefix naturally cluster in space; this prefix distance can
serve directly as the reward signal for RL alignment. Second, the geographic
hierarchy enables \emph{inference acceleration}: the user's 4-digit geohash
identifies an $\approx39\,\text{km}$ cell; together with its 8 surrounding cells
this 3$\times$3 grid essentially covers the platform's maximum delivery radius.
The first two geographic tokens are therefore constrained to these $\le9$
prefixes $\mathcal{N}_u$, reducing full 6-token generation to 4 free tokens;
only the third geographic token is generated freely to locate the specific
$\approx1.2\,\text{km}$ cell within the reachable neighborhood.

\subsection{Encoder--Decoder Architecture}

\subsubsection{Encoder}
In on-demand delivery, user behaviors span diverse contexts and regions, with preferences shifting across locations;
in search, the user's current query intent is especially important. Aggregating
historical behaviors directly with standard self-attention is easily dominated by
interactions that are frequent but irrelevant to the current query and the
user's geographic preference. We therefore design a two-level encoder: the bottom level models
the context of the behavior sequence at the token level with self-attention,
and the top level filters and aggregates at the field level using the current
request as the target.

\noindent\mbox{\textbf{Query-Geo Target Attention.}} Given the user location $p_u$, the
query $q$, and the behavior sequence $\mathcal{B}=\{b_i\}_{i=1}^{L_b}$ with
$b_i=(p_i^{u}, q_i,g_i,s_i)$, the raw tokens are concatenated and encoded by
$L_e$ layers of self-attention, then mean-pooled per field to obtain the
four-field representation $(\mathbf{p}_i^{u},\mathbf{q}_i,\mathbf{g}_i,\mathbf{s}_i)$
of behavior $i$, serving as the key/value source of the upper level attention. The
request-side real-time location representation $\mathbf{p}_u$ and query
representation $\mathbf{q}$ jointly form the target; the geographic path uses
$\mathbf{p}_u$ as query and $[\mathbf{p}_i^{u};\mathbf{g}_i]$ as key,
and the semantic path uses $\mathbf{q}$ as query and $[\mathbf{s}_i;\mathbf{q}_i]$
as key. The two paths are dynamically weighted by a gate
$\alpha$ and normalized:
\begin{equation}
e_i=\frac{1}{\sqrt{d}}\Big(\alpha\,\big\langle \mathbf{W}_g\mathbf{p}_u,\;
\mathbf{U}_g[\mathbf{p}_i^{u};\mathbf{g}_i]\big\rangle+(1-\alpha)\,
\big\langle \mathbf{W}_s\mathbf{q},\;\mathbf{U}_s[\mathbf{s}_i;\mathbf{q}_i]\big\rangle\Big),
\end{equation}
\begin{equation}
\mathbf{r}_{\text{int}}=\sum_{i=1}^{L_b}\mathrm{softmax}_i(e)\,\mathbf{v}_i,
\end{equation}
where $\alpha=\sigma(\mathbf{w}_\alpha^\top[\mathbf{p}_u;\mathbf{q}])$,
$\mathbf{v}_i$ is a linear projection of the concatenated four-field
representation, and $\mathbf{W}_g,\mathbf{W}_s,\mathbf{U}_g,\mathbf{U}_s$ are
learnable matrices. The geographic key carries the user location at the time of
each past interaction, so that interactions made far from the user's current
position are down-weighted; the semantic key fuses merchant semantic content and
past queries, so intent matching does not rely solely on the interacted merchant.

\noindent\mbox{\textbf{Query Fusion.}} For behavior-sparse users, $\mathbf{r}_{\text{int}}$
is noisy and prone to representation collapse, so we fuse it with the query
representation via a gate, falling back to $\mathbf{q}$ when the behavior signal
is weak:
\begin{equation}
\boldsymbol{\eta} = \sigma\big(\mathbf{W}_\eta[\mathbf{r}_{\text{int}};\mathbf{q}]\big),
\end{equation}
\begin{equation}
\mathbf{r}_u = \boldsymbol{\eta}\odot\mathbf{r}_{\text{int}} + (\mathbf{1}-\boldsymbol{\eta})\odot\mathbf{q},
\end{equation}
where $\mathbf{r}_u$ is the memory for the decoder cross-attention.
Compared with standard self-attention, this module
splits geographic and semantic similarity into two independent query/key
projections and explicitly decouples them via a cross-attention; the gate weight $\alpha$ is interpretable and adjustable. With a
lightweight structure, it simultaneously addresses cross-context and
cross-region interest transfer and provides a fallback for sparse users.

\subsubsection{Decoder}
We adopt an autoregressive \mbox{T5-style} Transformer decoder to generate the 6-token
Geo-Semantic ID $\hat{\mathbf{y}}=[\hat{y}_1,\dots,\hat{y}_6]$ step by step, with
the user representation $\mathbf{r}_u$ (Eq.~(8)) as cross-attention
memory. We decompose the generation at a single step $t$ into three stages:
input-side gated injection, per-layer attention, and output-side constrained
decoding.

\noindent\mbox{\textbf{User-Geo Gate.}} Since the merchant geographic tokens depend
only on the user's current geohash and are independent of the
semantic tokens, we inject the user's position-aligned location embedding
$\mathbf{p}^{(k)}$ only when decoding a geographic token
($t=k\in\{1,2,3\}$), and use a learnable gate $\boldsymbol{\gamma}_t$ to
modulate the injection strength per dimension; the semantic tokens ($t\in\{4,5,6\}$)
remain unchanged to avoid polluting content generation:
\begin{equation}
\begin{aligned}
&\boldsymbol{\gamma}_t=\sigma\!\big(\mathbf{W}_\gamma[\mathbf{h}_t^{(0)};\mathbf{p}^{(k)}]+\mathbf{b}_\gamma\big),\\
&\tilde{\mathbf{h}}_t^{(0)}=\begin{cases}
\mathbf{h}_t^{(0)}+\boldsymbol{\gamma}_t\odot\mathbf{p}^{(k)}, & t=k\in\{1,2,3\},\\[2pt]
\mathbf{h}_t^{(0)}, & t\in\{4,5,6\}.
\end{cases}
\end{aligned}
\end{equation}
The learned gate $\boldsymbol{\gamma}_t\in(0,1)^{d}$, with learnable
$\mathbf{W}_\gamma$ and $\mathbf{b}_\gamma$, is expected to approach $1$
when the location embedding is informative, and to decay when the
target lies in a neighboring cell, allowing cross-boundary exploration, thus
injecting proximity as a soft bias rather than imposing it as a hard constraint. The fused $\tilde{\mathbf{h}}_t^{(0)}$ is
fed into $L$ decoder blocks, each performing causal masked self-attention,
cross-attention to the memory $\mathbf{r}_u$, and a feed-forward network, all
with residuals and Pre-Norm (RMSNorm):
\begin{equation}
\begin{aligned}
\mathbf{a}^{(l)} &= \mathbf{h}^{(l-1)} + \text{MHA}_{\text{self}}\!\big(\text{Norm}(\mathbf{h}^{(l-1)}),\ \text{mask}_{\le t}\big),\\
\mathbf{c}^{(l)} &= \mathbf{a}^{(l)} + \text{MHA}_{\text{cross}}\!\big(\text{Norm}(\mathbf{a}^{(l)}),\ \mathbf{r}_u\big),\\
\mathbf{h}^{(l)} &= \mathbf{c}^{(l)} + \text{FFN}\!\big(\text{Norm}(\mathbf{c}^{(l)})\big),\quad l=1,\dots,L.
\end{aligned}
\end{equation}
On the output side, the final-layer output $\mathbf{h}_t^{(L)}$ is projected by
the output matrix to logits over the codebook of the current field, and a
softmax yields the token distribution from which beam search expands candidates:
\begin{equation}
P(\hat{y}_t\mid \hat{y}_{<t},\mathbf{r}_u)=\text{softmax}\!\big(\mathbf{W}_o\,\mathbf{h}_t^{(L)}\big).
\end{equation}
Decoding proceeds autoregressively for $t=1,\dots,6$, producing a valid \gsid{}
token by token; the beam retains the top-$B$ deduplicated sequences, forming the
retrieved candidate set for the request.

\subsubsection{Training Objective}
The encoder--decoder is pre-trained from scratch with next-token
prediction. Training samples are drawn from real click logs: the request context
$(q,p_u,\mathcal{B})$ is the input, and the \gsid{}
$\mathbf{y}^{\ast}=[y_1^{\ast},\dots,y_6^{\ast}]$ of the clicked merchant is the
target sequence, optimized token by token via teacher forcing:
\begin{equation}
\mathcal{L}_{\text{NTP}} = -\,\mathbb{E}_{(q,p_u,\mathcal{B},\mathbf{y}^{\ast})\sim\mathcal{D}}\ \sum_{t=1}^{6}\log P\big(y_t^{\ast}\mid \mathbf{y}_{<t}^{\ast},\,\mathbf{r}_u\big).
\end{equation}
The geographic and semantic tokens share a single cross-entropy;
position-specific weights can be assigned to emphasize certain token positions
based on business priorities, though we use uniform weights for simplicity. The
converged model is denoted $\pi_{\text{pre}}$, serving both as the initialization
of the RL in \S\ref{sec:gmpo} and as its KL-regularization reference model.

\subsection{GMPO Alignment}
\label{sec:gmpo}
The supervision above is token-level and fits only online
exposure data, but, constrained by exposure bias and this local objective,
it cannot capture group-level coverage or the business-value hierarchy of
retrieval. Therefore, in the alignment stage we introduce set-level
prediction: the whole deduplicated candidate set from beam search is served to
the user as a unit, and its business value is measured by a composite reward
over set-level coverage. For each query, the policy $\pi_\theta$ samples
$G$ \gsid{} sequences, retaining all sample indices, including repeated
merchant IDs, with $\pi_{\text{pre}}$ as the reference model. Each sampled
sequence $y^{(i)}$ identifies one merchant, represented by the singleton
$\mathcal{Y}^{(i)}=\{y^{(i)}\}$ for reward computation; membership in
$\mathcal{B}_c$ is evaluated using the corresponding merchant ID.
Compared with the pre-training stage, we augment the behavior data with order
sequences and high-ad-value sequences (filtered from online advertising logs),
denoted
$\mathcal{B}=\mathcal{B}_{\text{clk}}\cup\mathcal{B}_{\text{ctr}}\cup\mathcal{B}_{\text{ord}}$,
corresponding to click, high-CTR, and order sequences respectively, with the
three reward classes denoted $\mathcal{C}=\{\text{clk},\text{ctr},\text{ord}\}$. The
single-sample reward (SR) measures the weighted hit score of the $i$-th sampled
output $\mathcal{Y}^{(i)}$ against the behavior set $\mathcal{B}$:
\begin{equation}
\text{SR}^{(i)} = \sum_{y\in\mathcal{Y}^{(i)}}\sum_{c\in\mathcal{C}} w_c\cdot\mathbf{1}[y\in\mathcal{B}_c].
\end{equation}
The weights follow the advertising business-value hierarchy: order $>$ click $>$
high-CTR.

\noindent\mbox{\textbf{Group Reward (GR).}} Let
$\mathcal{Y}_g=(\mathcal{Y}^{(1)},\ldots,\mathcal{Y}^{(G)})$ denote the
indexed group of sampled outputs, not their deduplicated union. For any
sample group $\mathcal{S}$, define
$\operatorname{uniq}(\mathcal{S})=\bigcup_{\mathcal{Y}\in\mathcal{S}}\mathcal{Y}$
as its set of distinct merchant IDs, with
$\operatorname{uniq}(\varnothing)=\varnothing$ and $\text{GR}(\varnothing)=0$.
Deduplication occurs only when computing the group reward, so each merchant
contributes once:
\begin{equation}
\text{GR}(\mathcal{S}) = \sum_{y\in\operatorname{uniq}(\mathcal{S})}\sum_{c\in\mathcal{C}} w_c\cdot\mathbf{1}[y\in\mathcal{B}_c].
\end{equation}

\noindent\mbox{\textbf{Marginal Advantage (MA).}} Define
$\mathcal{Y}_g^{-i}=(\mathcal{Y}^{(j)})_{j\in\{1,\ldots,G\}\setminus\{i\}}$
by removing only the $i$-th sample from the indexed group, while retaining
all other samples, including those with the same merchant ID. We then
recompute the deduplicated coverage of the remaining group. The final
advantage combines the z-scored SR (standardized across the $G$ samples)
and this leave-one-sample-out contribution with $\lambda\in[0,1]$,
reducing to the standard GRPO advantage when $\lambda=1$; when
$\lambda=0$, only MA is used and the SR standardization term is omitted:
\begin{equation}
\begin{gathered}
\text{MA}^{(i)} = \text{GR}(\mathcal{Y}_g) - \text{GR}(\mathcal{Y}_g^{-i}), \\
A_i = \lambda\cdot\mathrm{zscore}(\text{SR}^{(i)}) + (1-\lambda)\cdot\text{MA}^{(i)}.
\end{gathered}
\end{equation}
For example, if two samples both identify merchant $a$, removing either
sample leaves $a$ covered by the other, so both receive $\text{MA}=0$.
With nonnegative reward weights, only newly covered rewarded merchants
receive positive marginal credit; a duplicate receives no positive
credit, rather than an explicit negative penalty. This discourages
redundant coverage but does not guarantee recovery from policy collapse:
if all sampled IDs are identical and $G>1$, all marginal advantages are zero.
The final training objective follows the GRPO formulation with importance ratio
$\rho_i=\pi_\theta(y^{(i)})/\pi_{\theta_{\text{old}}}(y^{(i)})$
and KL regularization to anchor the policy near the pre-trained reference:
\begin{equation}
\mathcal{L}(\theta) = -\mathbb{E}\!\left[\min\!\big(\rho_i A_i,\,\mathrm{clip}(\rho_i,1{\pm}\epsilon)\,A_i\big)\right] + \beta\,\mathrm{KL}(\pi_\theta\|\pi_{\text{pre}}).
\end{equation}
In this way the policy is optimized toward set-level business value while
staying close to the pre-trained distribution.

\section{Experiments}

\subsection{Experimental Setup}

\subsubsection{Datasets}
We evaluate on one industrial dataset and two public datasets
(Table~\ref{tab:datasets}). The industrial
dataset (\textbf{TS-Delivery}) is drawn from real user behavior
logs of Taobao Shangou search advertising, spanning 12 uniformly sampled days,
with the first 10 days used for training, the 11th for validation, and the 12th
for testing.
Each sample carries the query text, the user's real-time coordinates, and the
global behavior sequence, aligned one-to-one with the problem definition in
\S\ref{sec:prob}. The public datasets use the NYC and TKY
subsets of Foursquare check-in data. Following GNPR~\cite{gnprsid}, we preprocess the
data, organize each user's behavior sequence chronologically, and split the data
80\%/10\%/10\% for training/validation/testing. Note that since current public
POI data lack real search queries, we adopt the same non-search preprocessing
used for the industrial dataset and set the query to a default placeholder,
testing model generality under a unified protocol. Since the public datasets
provide only per-POI coordinates and no real-time user location, we take the
coordinates of each user's most recent check-in POI as a proxy for the user's
current location.

\begin{table}[t]
  \centering
  \caption{Dataset statistics.}
  \label{tab:datasets}
  \begin{tabular}{llrrr}
    \toprule
    Dataset & & \#Users & \#Items & \#Interactions \\
    \midrule
    TS-Delivery & & 0.1B & 4.8M & 0.5B \\
    \multirow{2}{*}{Foursquare} & NYC & 1083 & 5135 & 147938 \\
                                & TKY & 2293 & 7873 & 447570 \\
    \bottomrule
  \end{tabular}
\end{table}

\subsubsection{Evaluation Metrics}
\noindent\mbox{\textbf{HitRate@$N$@$\delta$ (HR).}} Hit rate over reachable
candidates. It simulates the recall ability of the online pipeline, where distance
filtering precedes top-$N$ truncation: following the beam order, keep only
reachable ($d\le\delta$) deduplicated merchants, retain up to $N$ merchants
to form the reachable candidate
list $\mathcal{L}_u^{\delta,N}$, and mark the sample as a hit if it intersects
the ground-truth clicked-merchant set $\mathcal{G}_u$:
\begin{equation}
\text{HitRate@}N\text{@}\delta = \tfrac{1}{|\mathcal{U}'|}\!\sum_{u\in\mathcal{U}'}\!\mathbf{1}\big[\mathcal{L}_{u}^{\delta,N}\cap\mathcal{G}_{u}\neq\varnothing\big].
\end{equation}
Here $\mathcal{U}'$ is the test set of samples having both a user location and a
clicked merchant, $\mathbf{1}[\cdot]$ is the indicator function, and $d$ is the
spherical distance between merchant and user.

\noindent\mbox{\textbf{ValidRate@$N$@$\delta$ (VR).}} Measures the
geographic validity of the raw retrieval set: take the top-$N$
deduplicated merchants of the candidate set (no distance filtering), compute the
fraction within $\delta$ of the user, and average over samples:
\begin{equation}
\text{ValidRate@}N\text{@}\delta = \tfrac{1}{|\mathcal{U}|}\!\sum_{u\in\mathcal{U}}\frac{\big|\{s\in\mathcal{S}_{u}^{1:N}: d(s,p_{u})\le\delta\}\big|}{|\mathcal{S}_{u}^{1:N}|}.
\end{equation}
Here $\mathcal{S}_u^{1:N}$ is the top-$N$ merchants of $\mathcal{S}_u$, and
$\mathcal{U}$ is the test set of samples with a user location. A higher value
means the model's retrieval results are inherently closer to the user's current
location.

\noindent\mbox{\textbf{Coverage$_c$@$W$ (average group coverage).}} For GMPO
evaluation, measures the average number of ground-truth hits within a
deduplicated group of generated sequences (beam or sampling, group width $W$)
against a class of ground-truth behaviors:
\begin{equation}
\text{Coverage}_c@W=\tfrac{1}{|\mathcal{U}|}\!\sum_{u\in\mathcal{U}}\big|\mathcal{Y}_u\cap\mathcal{B}_c^{u}\big|,
\end{equation}
where $c\in\{\text{clk},\text{ctr},\text{ord}\}$.

\subsubsection{Baselines}
\textbf{Dense Retrieval} uses a location-conditioned DSSM-style factorized
two-tower architecture~\cite{dssm} with query, historical behavior, and geographic
features, retrieving from the full merchant corpus before distance filtering.
\textbf{TIGER}~\cite{tiger} is a pioneering generative-retrieval
method that quantizes item content into hierarchical SIDs with RQ-VAE and
autoregressively generates target SIDs. Its identifiers carry no spatial
structure and thus represent the ``plain SID + post-hoc geo-filter''
approach. \textbf{GNPR}~\cite{gnprsid} builds content-semantic
hierarchical SIDs for POIs with a diversity objective and fuses geographic
information into the quantization input, a soft-modeling approach to spatial
awareness. \textbf{OneLoc}~\cite{oneloc} is a geo-aware generative
recommender for local-life services combining geo-aware SID, geo-aware
self-attention, and a geo/GMV dual-reward DPO; it is the closest baseline to our
work and represents the strongest soft geo-guidance scheme.

For comparability, TIGER, GNPR, and OneLoc use the same data, model size,
and beam configuration. TIGER retains its original location-agnostic inputs,
with the same online geographic post-filter applied to its output.

\begin{table*}[t]
  \centering
  \caption{Main results on the industrial dataset TS-Delivery. HR denotes
  HitRate@$N$@$\delta$. Best in bold. The superscript $^{*}$ indicates a
  statistically significant improvement over OneLoc ($p<0.05$).}
  \label{tab:main}
  \setlength{\tabcolsep}{5pt}
  \renewcommand{\arraystretch}{0.9}
  \begin{tabular}{l ccc ccc ccc}
    \toprule
    & \multicolumn{3}{c}{$\delta$=5\,km} & \multicolumn{3}{c}{$\delta$=10\,km} & \multicolumn{3}{c}{$\delta$=20\,km}\\
    \cmidrule(lr){2-4}\cmidrule(lr){5-7}\cmidrule(lr){8-10}
    Model & HR@50 & HR@150 & HR@300 & HR@50 & HR@150 & HR@300 & HR@50 & HR@150 & HR@300\\
    \midrule
    Dense Retrieval & 0.6710 & 0.7570 & 0.7930 & 0.6050 & 0.7240 & 0.7810 & 0.4980 & 0.6460 & 0.7160\\
    TIGER   & 0.7659 & 0.8512 & 0.8814 & 0.6959 & 0.8293 & 0.8858 & 0.5796 & 0.7464 & 0.8257\\
    GNPR    & 0.7678 & 0.8526 & 0.8849 & 0.6986 & 0.8319 & 0.8893 & 0.5834 & 0.7501 & 0.8283\\
    OneLoc  & 0.7778 & 0.8582 & 0.8844 & 0.7092 & 0.8410 & 0.8945 & 0.5929 & 0.7596 & 0.8370\\
    \method & \textbf{0.7869}$^{*}$ & \textbf{0.8593}$^{*}$ & \textbf{0.8876}$^{*}$ & \textbf{0.8090}$^{*}$ & \textbf{0.8925}$^{*}$ & \textbf{0.9313}$^{*}$ & \textbf{0.8080}$^{*}$ & \textbf{0.8920}$^{*}$ & \textbf{0.9336}$^{*}$\\
    \bottomrule
  \end{tabular}
\end{table*}

\subsubsection{Implementation Details}
We train with the AdamW optimizer on 20 PPU-ZW810E GPUs, using an initial
learning rate of $1.5\times10^{-4}$ and a weight decay of 0.1. The quantization
module comprises six codebook levels: the geographic codebook has $K_g=1024$
geohash cells per level across $L_g=3$ levels, and the content codebook has
$K_c=256$ clusters per level across $L_c=3$ levels, so each merchant is
represented by 6 tokens (3 geographic + 3 content). The encoder stacks 2
token-level Transformer blocks and one dual-path target-attention module, and the
decoder stacks 12 blocks, with a hidden dimension of 512, 8 attention heads, and
a feed-forward (FFN) dimension of 2048. The user historical behavior sequence
length $L_b$ is set to 100. Offline evaluation and online serving both decode with a beam width of 300,
from which the top-$N$ deduplicated candidates are taken.
In the RL stage, the group size $G$ is set to 8, the
advantage weight $\lambda$ to 0 (so the deployed advantage is the pure
marginal advantage), the KL coefficient $\beta$ to 0.01, and the clip
threshold $\epsilon$ to 0.2; the per-behavior reward weights
$w_{\text{clk}}$, $w_{\text{ctr}}$, and $w_{\text{ord}}$ are set to 1.0, 0.1, and
5.0, respectively. These weights are not tuned on the test set but follow the
business-value hierarchy of on-demand delivery: an order is worth far more than a
click, and a click more than an impression-level CTR signal.

\subsection{Experimental Results}

\subsubsection{Industrial Dataset}
Table~\ref{tab:main} summarizes results on TS-Delivery under three delivery radii
($\delta\in\{5,10,20\}$\,km) and three candidate depths ($N\in\{50,150,300\}$).
Note that in production, candidates are retrieved within a 20\,km radius.
\method{} also outperforms Dense Retrieval at all nine radius--depth settings.

Reading the table across radii, we observe that the gap between \method{} and
the generative baselines is small at $\delta=5$\,km but widens sharply as $\delta$ grows: on
HR@50 the improvement over the strongest baseline OneLoc is 1.2\% at 5\,km
(0.7778$\to$0.7869), 14.1\% at 10\,km (0.7092$\to$0.8090), and 36.3\% at 20\,km
(0.5929$\to$0.8080). This pattern correlates with the size of the reachable set. At a
small $\delta$ the set is limited, so the first 50 reachable merchants already cover
much of it and all generative methods approach saturation; as $\delta$ grows the reachable
set expands, and the task shifts toward selecting, within this larger set,
the nearby merchants a user is likely to click. Because \method{} constrains
generation to the user's reachable neighborhood (\S\ref{sec:reach}), its
candidates stay concentrated in this region, which is where the larger 10\,km and
20\,km margins come from. \method{} still takes the best HR under every $N$--$\delta$
combination; at the operational setting ($\delta=20$\,km, the production
retrieval radius) the gains over OneLoc reach 36.3\%/17.4\%/11.5\% on HR@50/150/300.

We also observe that HR@300 decreases for TIGER, GNPR, and OneLoc when the
distance cutoff increases from 10 to 20\,km. The beam width of 300 caps
predicted SIDs, not mapped merchants; one-to-many mapping can yield more
than 300 candidates. Since the same mapped pool is distance-filtered before
top-$N$ truncation, the larger cutoff admits high-ranked merchants
10--20\,km away, potentially displacing clicked merchants within 10\,km
from the final top-300 list. HR@300 therefore need not be monotonic in
$\delta$.

\subsubsection{Public Datasets}
Table~\ref{tab:public} summarizes hit-rate results on the Foursquare dataset
under three candidate depths ($N\in\{5,10,20\}$).

\begin{table}[t]
  \centering
  \caption{Results on the public Foursquare dataset. Best in bold.
  The superscript $^{*}$ indicates a statistically significant improvement
  over OneLoc ($p<0.05$).}
  \label{tab:public}
  \setlength{\tabcolsep}{4pt}
  \resizebox{\columnwidth}{!}{%
  \begin{tabular}{l ccc ccc}
    \toprule
    & \multicolumn{3}{c}{NYC} & \multicolumn{3}{c}{TKY}\\
    \cmidrule(lr){2-4}\cmidrule(lr){5-7}
    Model & HR@5 & HR@10 & HR@20 & HR@5 & HR@10 & HR@20\\
    \midrule
    TIGER   & 0.5146 & 0.6288 & 0.7189 & 0.4658 & 0.5623 & 0.6313\\
    GNPR    & 0.5279 & 0.6431 & 0.7345 & 0.4978 & 0.5991 & 0.6716\\
    OneLoc  & 0.6238 & 0.7246 & 0.7868 & 0.5186 & 0.6116 & 0.6812\\
    \method & \textbf{0.6464}$^{*}$ & \textbf{0.7472}$^{*}$ & \textbf{0.8246}$^{*}$ & \textbf{0.5388}$^{*}$ & \textbf{0.6414}$^{*}$ & \textbf{0.7218}$^{*}$\\
    \bottomrule
  \end{tabular}}
\end{table}

Across both public subsets we observe that \method{} again ranks first on every
metric, improving over OneLoc by 3.1\%--4.8\% on NYC and 3.9\%--6.0\% on
TKY. The ordering TIGER $<$ GNPR $<$ OneLoc $<$ \method{} reproduces the
industrial result on a very different data distribution, which suggests that the
link between stronger geographic modeling and better recall is not an artifact of
one dataset. The margin here ($\approx$3\%--6\%) is much smaller than the
14.1\%--36.3\% seen at 10\,km/20\,km on TS-Delivery; this is expected, since
Foursquare has no delivery radius and $N$ only reaches 20, so the structural
reachability constraint is largely inactive, and the placeholder query leaves no
query--geo signal to exploit. The remaining gain therefore comes from jointly
modeling user interest and POI geography. The TKY numbers
sit below NYC throughout, which we attribute to its larger item set
(7.9K vs.\ 5.1K), denser per-user interactions, and more concentrated POI
distribution.

\subsubsection{Ablation Study}
Table~\ref{tab:ablation} summarizes the ablation results of each sub-module on
TS-Delivery.

\begin{table*}[t]
  \centering
  \caption{Ablation study on TS-Delivery. HR denotes HitRate@$N$@$\delta$.
  The full model is in bold.}
  \label{tab:ablation}
  \setlength{\tabcolsep}{5pt}
  \begin{tabular}{l ccc ccc ccc}
    \toprule
    & \multicolumn{3}{c}{$\delta$=5\,km} & \multicolumn{3}{c}{$\delta$=10\,km} & \multicolumn{3}{c}{$\delta$=20\,km}\\
    \cmidrule(lr){2-4}\cmidrule(lr){5-7}\cmidrule(lr){8-10}
    Model & HR@50 & HR@150 & HR@300 & HR@50 & HR@150 & HR@300 & HR@50 & HR@150 & HR@300\\
    \midrule
    w/o geo prefix \& u-geo gate & 0.7819 & 0.8614 & 0.8860 & 0.7124 & 0.8451 & 0.8989 & 0.5960 & 0.7632 & 0.8403\\
    w/o geo prefix & 0.7794 & 0.8580 & 0.8835 & 0.7111 & 0.8416 & 0.8945 & 0.5947 & 0.7606 & 0.8371\\
    w/o u-geo gate & 0.7773 & 0.8497 & 0.8775 & 0.7961 & 0.8807 & 0.9203 & 0.7934 & 0.8798 & 0.9224\\
    w/o query-geo attention & 0.7661 & 0.8462 & 0.8789 & 0.7836 & 0.8750 & 0.9211 & 0.7820 & 0.8735 & 0.9230\\
    w/o query fusion & 0.7865 & 0.8588 & 0.8871 & 0.8092 & 0.8922 & 0.9308 & 0.8080 & 0.8919 & 0.9331\\
    full model & \textbf{0.7869} & \textbf{0.8593} & \textbf{0.8876} & \textbf{0.8090} & \textbf{0.8925} & \textbf{0.9313} & \textbf{0.8080} & \textbf{0.8920} & \textbf{0.9336}\\
    \bottomrule
  \end{tabular}
\end{table*}

We analyze Table~\ref{tab:ablation} one component at a time. Removing the geo prefix
(rows 1--2) pulls HR@50 at 10\,km and 20\,km back to baseline levels (0.7111 vs.\
0.7092 for OneLoc at 10\,km; 0.5947 vs.\ 0.5929 at 20\,km), while the 5\,km numbers
barely move. Without the structural constraint the generated candidates spread
across the whole ID space; at narrow radii the post-hoc filter still recovers
most hits, but at wide radii the lost geographic concentration leaves the model on
par with a pure-semantic baseline, so the geo prefix accounts for most of the
large-radius gain (\S\ref{sec:reach}). Comparing the two rows also shows that the
prefix and the gate act jointly rather than additively: once the prefix is gone,
row~1 is uniformly better than row~2, because a gate conditioned on the user's
real-time location can only reorder cells that the prefix has already restricted
to the user's geographic neighborhood; without that restriction it reweights an unconstrained ID space
and merely injects noise into the decoding order.

Dropping the User-Geo Gate alone (row 3) lowers HR@50 by 0.96/1.29/1.46
percentage points across the three radii, and the damage grows with the radius as
the reachable set expands. We interpret this as a division of labor: the prefix
code fixes which geographic tokens are admissible, whereas the gate, conditioned
on the user's real-time location, decides which reachable cell to generate first.

The two interest-modeling components behave very differently. Removing Query-Geo
Target Attention (row 4) reduces HR@50 by 2.08/2.54/2.60 percentage points across
the three radii. The drop is larger than for the User-Geo Gate and grows more
slowly with the radius in relative terms. This relative flatness suggests that it works on the interest
representation rather than on reachability: whatever the size of the reachable
set, aligning the attention target with the current query and location helps
the encoder pick the merchant the user actually wants, so its benefit does not fade
at small radii the way the geo prefix does. Query fusion (row 5) is by contrast
almost invisible on the full population (within 0.05 percentage points); it is a fallback
path for behavior-sparse users, whose effect is diluted over all traffic and
should matter more on the sparse-user segment.

\subsubsection{Geographic Validity}
\label{sec:reach}
\begin{figure}[t]
  \centering
  \includegraphics[width=\columnwidth]{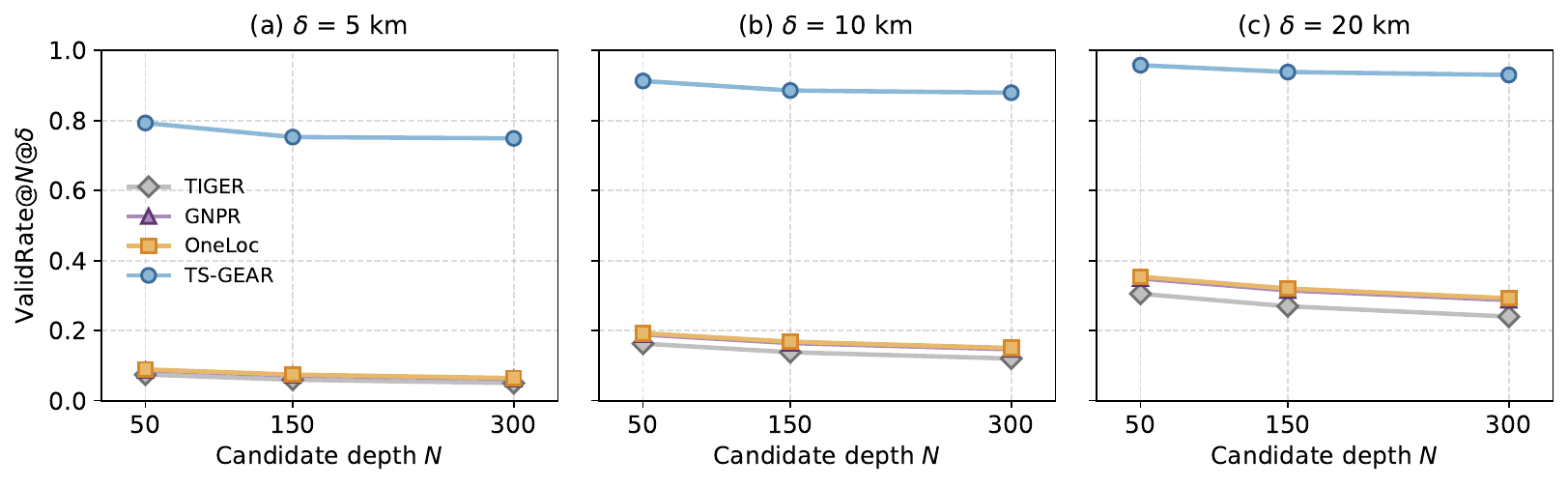}
  \caption{Geographic validity (ValidRate@$N$@$\delta$) across
  candidate depths. \method{} keeps validity roughly in the 75\%--96\% range
  across depths and radii, whereas baselines remain around 35\% or lower.}
  \label{fig:reach}
\end{figure}

We separately report VR@$N$@$\delta$, the fraction of generated candidates
within the delivery radius, for all models (Figure~\ref{fig:reach}).

We center the analysis on $\delta=20$\,km, the radius used in production. There
\method{} keeps 95.8\% of its top-50 candidates within the delivery radius,
against 35.4\% for the strongest baseline OneLoc, and the margin persists as the
beam deepens: 93.9\% at $N=150$ and 93.0\% at $N=300$, versus 32.0\% and 29.3\%.
Validity is uniformly high at the smaller radii as well: \method{} never drops
below 74.9\% at any depth or radius, while every baseline stays at or below
35.4\% (Figure~\ref{fig:reach}).

Placed next to Table~\ref{tab:main}, these numbers show why validity and hit rate
do not move in lockstep. At 20\,km the reachable set is large, and the 95.8\%
validity of \method{} is what turns into the 36.3\% HR@50 gain, since keeping
most candidates within the delivery radius leaves fewer slots spent on
unreachable merchants. At smaller radii the reachable set saturates quickly, so
the same validity advantage yields only a 1.2\% HR gain at 5\,km.

We note that reachability is a necessary condition, not the objective: what
ultimately matters is the ranking \emph{within} the reachable set. A post-hoc
filter, or an inference-time distance mask over the unreachable candidates decoded from plain SIDs,
could also push validity toward 100\%, but such masking only renormalizes a
distribution trained without geography: the surviving beam stays reachable, yet its ordering
is location-agnostic, so validity rises while HR does not follow. Because
geography is instead generated end-to-end, the model learns a
location-conditioned ordering within the reachable set. The User-Geo Gate makes this concrete: removing it still
costs 0.96--1.46 percentage points of HR@50 even though the nine-cell prefix
constraint of \S\ref{sec:gsid} is left untouched. The gain we care about thus comes less from being reachable and
more from being correctly ordered inside the reachable neighborhood.
Although the prefix distance could also serve as a reward signal for RL
(\S\ref{sec:gsid}), the structural prefix constraint already yields over 95\%
validity, making a separate geographic reward
unnecessary; GMPO therefore focuses exclusively on business-value coverage.

\subsubsection{RL Alignment Analysis}
\begin{figure}[t]
  \centering
  \includegraphics[width=\columnwidth]{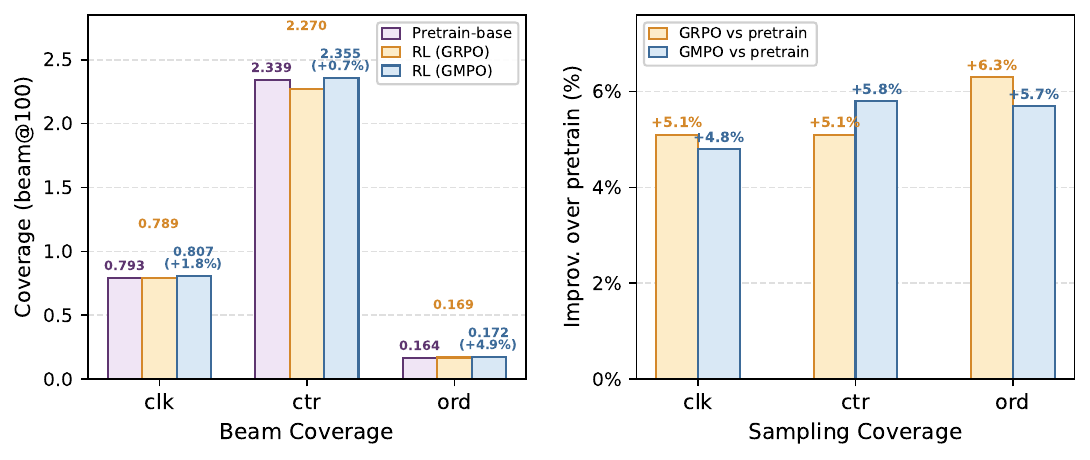}
  \caption{RL alignment analysis. (a) Cumulative coverage under the
  beam@100 protocol: standard GRPO declines on clk/ctr coverage, while GMPO
  improves all three signals (relative gains over pretrain-base in parentheses).
  (b) Relative improvement over pretrain-base under the sampling protocol: GRPO and
  GMPO are close and both clearly outperform pretrain-base, masking the policy
  collapse that the beam protocol exposes.}
  \label{fig:gmpo}
\end{figure}

Under the production decoding setting (beam@100 with deduplication), we compare
the coverage of the pre-trained baseline, standard GRPO, and GMPO
(Figure~\ref{fig:gmpo}(a)).

Here we observe that GRPO underperforms the pre-trained baseline on coverage
($-0.6\%$ on clk, $-2.9\%$ on ctr), with its only gain on ord ($+3.3\%$). The cause lies in the
data distribution of on-demand delivery search: up to 70\% of queries yield only
a single click, so pre-training and RL samples overlap heavily and the initial
policy is highly concentrated. Once all beam outputs collapse to the same
\gsid{}, the importance ratios become identical and the z-score-normalized
advantage sums to zero, so the gradient vanishes and remains zero indefinitely;
the policy stays locked onto a few high-reward modes. The ord reward, carrying the highest
weight, still improves, but at the expense of within-beam diversity and overall
coverage. In effect GRPO optimizes the expected reward of an individual sample,
which is misaligned with retrieval, where the unit of value is a set.

GMPO breaks this deadlock, raising clk/ctr/ord by
1.8\%/\allowbreak 0.7\%/\allowbreak 4.9\% over the pre-trained baseline and by
2.4\%/\allowbreak 3.7\%/\allowbreak 1.6\% over GRPO. The gains are largest on ord, in line with
$w_{\text{ord}}=5.0$; the uniform improvement across the three signals shows
that the marginal advantage restores within-beam diversity, while the ordering
of the gains still follows the business-value hierarchy. The clearest sign of this shift is ctr, which
moves from $-2.9\%$ to $+0.7\%$: high-CTR merchants that were shown but never
clicked are exactly the class a collapsed policy crowds out first.

GRPO and GMPO are nearly indistinguishable under the sampling protocol, where
both clearly outperform pretrain-base (Figure~\ref{fig:gmpo}(b)), and they
separate only under beam search. We attribute this to the randomness injected by
the sampling protocol, which sustains within-beam diversity and thereby masks the
policy collapse, whereas beam search deterministically takes the top-$W$ and
exposes the collapse. Since production serving decodes with beam search, judging
RL gains from the sampling protocol alone would overstate the gains realized
online.

\subsubsection{Scaling-Law Study}
\begin{figure}[t]
  \centering
  \includegraphics[width=\columnwidth]{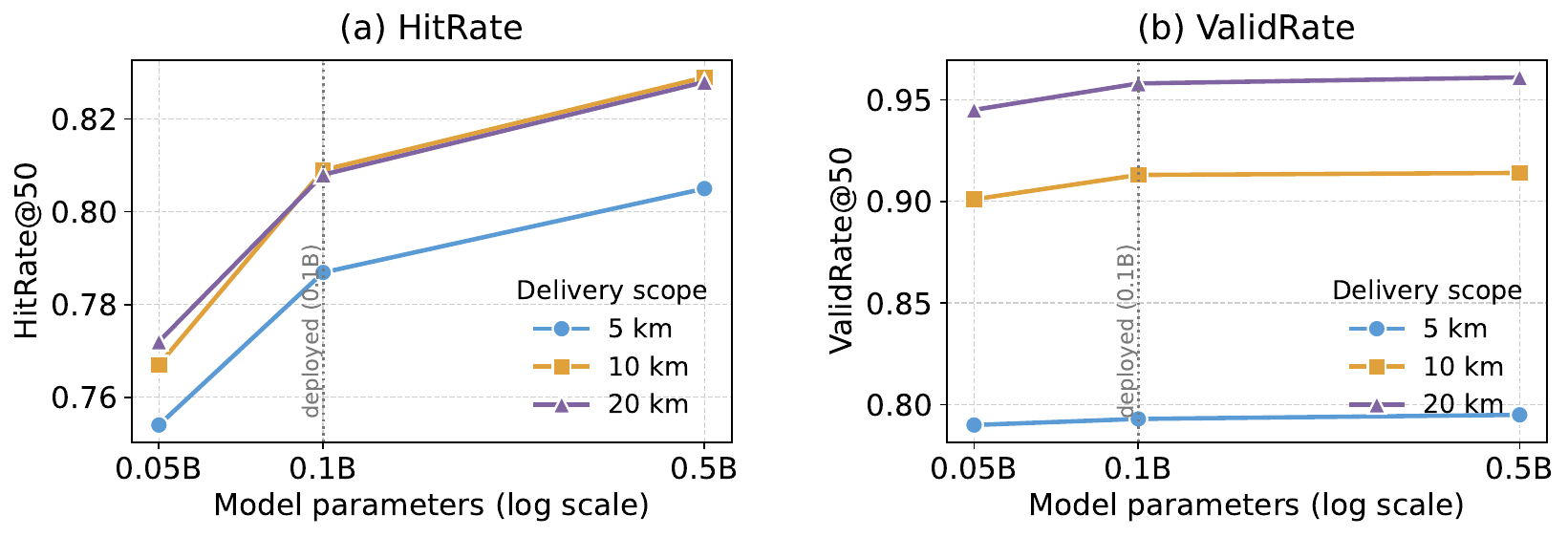}
  \caption{Scaling-law study across 0.05B / 0.1B / 0.5B backbones. HitRate
  scales roughly linearly with capacity, while geographic validity stays
  comparatively flat.}
  \label{fig:scaling}
\end{figure}

We vary the Transformer hidden size together with the number of encoder and
decoder layers, comparing 0.05B / 0.1B / 0.5B backbones under the same data and
training configuration (Figure~\ref{fig:scaling}). We observe that HitRate rises roughly
linearly with capacity, whereas geographic validity stays comparatively flat; the
latter is determined primarily by the prefix constraint rather than by model size. Under
the online latency constraint we deploy the 0.1B version.

\subsection{Online A/B Test}
\method{} is deployed on NVIDIA A10 GPUs (latency budget 70\,ms) with TensorRT,
mixed precision, and geographic-prefix-based decoding acceleration: given the
user's real-time location, a nine-cell lookup determines $\le9$ reachable geographic
prefixes as a forced prefill, skipping the first two geo-tokens and reusing the KV
cache. The module serves as one incremental recall path in parallel with existing
multi-channel retrieval (keyword, vector, etc.), with results fed jointly into
advertising ranking and bidding.

On the main Taobao Shangou search advertising pipeline, we ran a continuous 7-day
online experiment serving 20\% of traffic. Adding \method{} (0.1B backbone) to the
existing multichannel retrieval system, including keyword matching and two-tower
retrieval, etc., yielded system-level gains of 3.8\% in ad impressions, 0.4\%
in ad efficiency, and 4.2\% in ad revenue (Table~\ref{tab:online}; all $p<0.01$
under two-tailed t-tests).
\method{} is now fully deployed in production, serving over 150M search requests
per day.

\begin{table}[t]
  \centering
  \caption{Online A/B test results of \method{} on the Taobao Shangou search
  advertising platform (7-day online experiment serving 20\% of traffic; all metrics
  statistically significant with $p<0.01$ under two-tailed t-tests).}
  \label{tab:online}
  \begin{tabular}{lc}
    \toprule
    Online Metric & Improvement \\
    \midrule
    Ad Impressions & +3.8\% \\
    Ad Efficiency  & +0.4\% \\
    Ad Revenue     & +4.2\% \\
    \bottomrule
  \end{tabular}
\end{table}

\section{Conclusion}
This paper proposes \method{}, a generative retrieval framework for on-demand
delivery that addresses three key challenges: geographic hard
constraints, geographic-interest modeling, and set-level value alignment.
Geo-Semantic ID encodes geography as a prefix code so that the vast majority of
generated candidates are naturally within the delivery radius; Query-Geo Target
Attention conditions behavior aggregation on the current query and location,
enabling cross-region interest transfer;
GMPO aligns the policy with business value via group reward and marginal
advantage while suppressing policy collapse. Extensive offline and online
experiments demonstrate the effectiveness of these designs, and \method{} is now
fully deployed in production.
In the future, we plan to enrich the reward structure of GMPO with
incrementality and relevance rewards, and to advance the engineering of
larger-scale models together with the inference acceleration required to serve
them.

\section*{Ethical Considerations}
\noindent\mbox{\textbf{Location privacy.}}
\method{} conditions retrieval on the user's real-time coordinates, one of the
most sensitive signals a search system consumes. Location enters the model only
as geohash prefixes whose finest cell spans roughly
$1.2\,\text{km}\times0.6\,\text{km}$, so the model reads and generates coarse
regions rather than points, and the generated \gsid{}s identify merchants rather
than users. Training uses historical logs collected under the platform's
existing user agreement. Coarse quantization is not
anonymization, however: repeated cell-level observations across sessions can still
expose where a user lives or works, so access control on the request-time
location, retention limits on logged coordinates, and opt-out of location
personalization remain necessary.

\noindent\mbox{\textbf{Merchant fairness and fulfillment load.}}
Because decoding is restricted to the nine reachable cells and the reward is
dominated by business value, exposure concentrates on nearby, historically
high-converting merchants. Newly onboarded merchants and merchants in sparsely covered areas
therefore appear less often in the generated candidate set and must spend more
on advertising to obtain comparable exposure. The geographic prefix
constrains reachability rather than popularity, and the module is one incremental path
among several retrieval channels, so exposure floors, ad load caps (limiting the
fraction of ad slots in overall traffic), and concentration monitoring at the
ranking stage remain necessary complements. The same concentration has an
operational side: higher order volume becomes courier workload in the peak hours
and dense areas that the constraint favors, so retrieval gains should be reviewed
alongside dispatch capacity.


\end{document}